\documentstyle[prd,aps,floats,amsfonts,amssymb,amsmath]{revtex}

\begin{document}
 \renewcommand{\theequation}{\thesection.\arabic{equation}}
 \draft
  \title{Neutrino Optics and Oscillations in Gravitational Fields}
\author{G. Lambiase$^{a,b}$\thanks{E-mail: lambiase@sa.infn.it}, G. Papini$^{c,d}$\thanks{E-mail:
papini@uregina.ca}, R. Punzi$^{a, b}$\thanks{E-mail:
punzi@sa.infn.it}, G. Scarpetta$^{a,b,d}$\thanks{E-mail:
scarpetta@sa.infn.it}}
\address{$^a$Dipartimento di Fisica "E.R. Caianiello"
 Universit\'a di Salerno, 84081 Baronissi (SA), Italy.}
  \address{$^b$INFN - Gruppo Collegato di Salerno, Italy.}
  \address{$^c$Department of Physics, University of Regina, Regina, Sask, S4S 0A2, Canada.}
  \address{$^d$International Institute for Advanced Scientific Studies, 89019 Vietri sul Mare (SA), Italy.}
\date{\today}
\maketitle
\begin{abstract}

We study the propagation of neutrinos in gravitational fields
using wave functions that are exact to first order in the metric
deviation. For illustrative purposes, the geometrical background
is represented by the Lense-Thirring metric. We derive explicit
expressions for neutrino deflection, helicity transitions, flavor
oscillations and oscillation Hamiltonian.

\end{abstract}

\section{Introduction}
\setcounter{equation}{0}

Because of their coexistence in a host of problems, the
interaction of neutrinos with gravitational fields is of interest
in astrophysics and cosmology. At the same time, neutrino
detectors of increasing sensitivity and scope require a better
understanding of propagation and oscillation properties of
neutrinos in the neighborhood of massive astrophysical objects.

While in a few instances the treatment of gravity may necessitate
the complete apparatus of general relativity, using a weak field
approximation is sufficient in most problems that deal with
neutrinos. Even in the case of black holes, a first order
approximation in the metric deviation may still be adequate up to
distances of a few Schwarzschild radii from the gravitational
source. In the majority of instances, gravitation is treated as a
quasi-classical external field. Even so, the calculations tend to
be complex.

Useful, simplified treatments of the problem of neutrinos in
gravitational fields can be found in the literature, notably in
the heuristic approach of Cardall and Fuller \cite{CAR} in which
the authors construct an effective mechanical four-momentum
operator that incorporates spin and matter effects.

A more direct approach would require solving the Dirac equation
completely which is in general very difficult. It was however
shown in \cite{caipap,dinesh} that the covariant Dirac equation
can be solved exactly to first order in the metric deviation. This
procedure is manifestly covariant and gauge-invariant and only
requires the evaluation of path integrals that are trivial for
most physically relevant metrics. It can also be extended to
include electromagnetic fields and the effect of media on the
propagation of the particles. These are all issues of principle
one meets in the interaction of quantum systems with semiclassical
gravitational fields \cite{papini1}. They have been raised time
and again in the literature \cite{ahl}. We think it advantageous
to address them within the context of an approach that is
consistent and reproduces all gravitational effects that are
known, or are supposed to apply to fermions
\cite{hehlni,dinesh,lamb}. We propose to apply this method to the
study of neutrino optics, helicity transitions and flavor
oscillations.

In Sec. II we briefly review the essential points of the solution
found in \cite{caipap}. The neutrino geometrical optics is given
in Sec. III and applied in particular to the propagation of
neutrinos in a Lense-Thirring background for the distinct cases of
propagation parallel and orthogonal to the angular velocity of the
gravitational source. Helicity transitions and flavor oscillations
in the absence of matter are dealt with in Sec. IV and Sec. V for
the same two directions of propagation. Sec. VI contains the
conclusions.

\section{The Covariant Dirac equation}
\setcounter{equation}{0}

The behavior of spin-1/2 particles in the presence of a
gravitational field $g_{\mu\nu}$ is determined by the covariant
Dirac equation
\begin{equation}\label{DiracEquation}
  [i\gamma^\mu(x){\cal D}_\mu-m]\Psi(x)=0\,,
\end{equation}
where ${\cal D}_\mu=\nabla_\mu+i\Gamma_\mu (x)$, $\nabla_\mu$ is
the covariant derivative, $\Gamma_{\mu}(x)$ the spin connection
and the matrices $\gamma^{\mu}(x)$ satisfy the relations
$\{\gamma^\mu(x), \gamma^\nu(x)\}=2g^{\mu\nu}$. Both
$\Gamma_\mu(x)$ and $\gamma^\mu(x)$ can be obtained from the usual
constant Dirac matrices by using the vierbein fields $e_{\hat
\alpha}^\mu$ and the relations
\begin{equation}\label{II.2}
  \gamma^\mu(x)=e^\mu_{\hat \alpha}(x) \gamma^{\hat
  \alpha}\,,\qquad
  \Gamma_\mu(x)=-\frac{1}{4} \sigma^{{\hat \alpha}{\hat \beta}}
  e^\nu_{\hat \alpha}e_{\nu\hat{\beta};\, \mu}\,,
\end{equation}
where $\sigma^{{\hat \alpha}{\hat \beta}}=\frac{i}{2}[\gamma^{\hat
\alpha}, \gamma^{\hat \beta}]$. A semicolon and a comma are
frequently used as alternative ways to indicate covariant and
partial derivatives respectively. We use units $ \hbar = c = 1$
throughout the paper.

Eq. (\ref{DiracEquation}) can be solved exactly to first order in
the metric deviation
$\gamma_{\mu\nu}(x)=g_{\mu\nu}-\eta_{\mu\nu}$, where the Minkowski
metric $\eta_{\mu\nu}$ has signature -2. This is achieved by first
transforming (\ref{DiracEquation}) into the equation
\begin{equation}\label{DiracEqTrasf}
  [i{\tilde \gamma}^\nu (x) \nabla_{\nu}-m]{\tilde \Psi}(x)=0\,,
\end{equation}
where
\begin{equation}\label{PsiTilde}
{\tilde \Psi}(x)=S^{-1}\Psi(x)\,,\qquad S(x)=e^{-i\Phi_s(x)}\,,
\qquad \Phi_s(x)={\cal P}\int_P^x dz^\lambda \Gamma_\lambda (z)\,,
\qquad {\tilde \gamma}^{\mu}(x)=S^{-1}\gamma^{\mu}(x) S\,.
\end{equation}
By multiplying (\ref{DiracEqTrasf}) on the left by $(-i{\tilde
\gamma}^\nu (x)\nabla_{\nu}-m)$, we obtain the equation
\begin{equation}\label{KGequation}
  (g^{\mu\nu}\nabla_\mu\nabla_\nu+m^2){\tilde \Psi}(x)=0\,,
\end{equation}
whose solution
\begin{equation}\label{ExactSolution}
  {\tilde \Psi}(x)=e^{-i\Phi_G(x)}\Psi_0(x)\,,
\end{equation}
is exact to first order. The operator $\hat{\Phi}_G(x)$ is defined
as
\begin{equation}\label{PhiG}
  \hat{\Phi}_{G}=-\frac{1}{4}\int_P^xdz^\lambda\left[\gamma_{\alpha\lambda,
  \beta}(z)-\gamma_{\beta\lambda, \alpha}(z)\right]\hat{L}^{\alpha\beta}(z)+
  \frac{1}{2}\int_P^x dz^\lambda\gamma_{\alpha\lambda}\hat{k}^\alpha\,,
\end{equation}
\[
 [\hat{L}^{\alpha\beta}(z), \Psi_0(x)]=\left((x^\alpha-z^\alpha)\hat{k}^\beta-
 (x^\beta-z^\beta)\hat{k}^\alpha\right)\Psi_0(x)\,, \qquad
 [\hat{k}^\alpha, \Psi_0(x)]=i\partial^\alpha\Psi_0\,,
 \]
and $\Psi_0(x)$ satisfies the usual flat spacetime Dirac equation.
$\hat{L}_{\alpha\beta}$ and $\hat{k}^\alpha$ are the angular and
linear momentum operators of the particle. It follows from
(\ref{ExactSolution}) and (\ref{PsiTilde}) that the solution of
(\ref{DiracEquation}) can be written in the form
\begin{equation}\label{PsiSolution}
  \Psi(x)=e^{-i\Phi_s}\left(-i{\tilde \gamma}^\mu(x)\nabla_\mu
  -m\right)e^{-i\Phi_G}\, \Psi_0(x)\,,
\end{equation}
and also as
\begin{equation}\label{PsiSolution2}
  \Psi(x)=-\frac{1}{2m}\left(-i\gamma^\mu(x){\cal
  D}_\mu-m\right)e^{-i\Phi_T}\Psi_0(x)\,,
\end{equation}
where $\Phi_T=\Phi_s+\Phi_G$ is of first order in
$\gamma_{\alpha\beta}(x)$. The factor $ -1/2m $ on the r.h.s. of
(\ref{PsiSolution2}) is required by the condition that both sides
of the equation agree when the gravitational field vanishes.

It is useful to re-derive some known results from the covariant
Dirac equation. On multiplying (\ref{DiracEquation}) on the left
by $(-i\gamma^\nu(x){\cal D}_\nu-m)$ and using the relations
\begin{equation}\label{relation1}
  \nabla_\mu\Gamma_\nu(x)-\nabla_\nu\Gamma_\mu(x)+i[\Gamma_\mu(x),
  \Gamma_\nu(x)]=-\frac{1}{4}\sigma^{\alpha\beta}(x)R_{\alpha\beta\mu\nu}\,,
\end{equation}
and
\begin{equation}\label{relation2}
  [{\cal D}_\mu, {\cal D}_\nu]=-\frac{i}{4}\,
  \sigma^{\alpha\beta}(x)R_{\alpha\beta\mu\nu}\,,
\end{equation}
we obtain the equation
\begin{equation}\label{KGEq+R}
  \left(g^{\mu\nu}{\cal D}_\mu{\cal
  D}_\nu-\frac{R}{4}+m^2\right)\Psi(x)=0\,.
\end{equation}
In (\ref{relation2}) and (\ref{KGEq+R}) $R_{\alpha\beta\mu\nu}$ is
the Riemann tensor, $R$ the Ricci scalar, and
$\sigma^{\alpha\beta}(x)=(i/2)[\gamma^\alpha(x),
\gamma^\beta(x)]$.

By using Eq. (\ref{PsiTilde}), we also find
\begin{equation}\label{Solution}
  (-i\gamma^\nu(x){\cal D}_\nu-m)\,S\, (i{\tilde
  \gamma}^\mu\nabla_\mu-m){\tilde \Psi}(x)=
  S\, (g^{\mu\nu}\nabla_\mu\nabla_\nu+m^2){\tilde \Psi}(x)=0\,.
\end{equation}
While Eq. (\ref{Solution}) is mainly a re-statement of the fact
that (\ref{PsiSolution}) is a solution of (\ref{DiracEquation}),
Eq. (\ref{KGEq+R}) implies that the gyro-gravitational ratio of a
massive Dirac particle is one when $R\neq 0$, as found by Oliveira
and Tiomno \cite{oliveira}, Audretsch \cite{audretsch} and
Kannenberg \cite{kannenberg}.

It is known that the weak field approximation $
g_{\mu\nu}=\eta_{\mu\nu}+ \gamma_{\mu\nu}$ does not fix the
reference frame completely. The transformations of coordinates $
x_{\mu}\rightarrow x_{\mu}+\xi_{\mu}$, with $ \xi_{\mu}(x) $ also
small of first order, are still allowed and lead to the "gauge"
transformations $ \gamma_{\mu\nu}\rightarrow
\gamma_{\mu\nu}-\xi_{\mu,\nu}-\xi_{\nu,\mu}$. It is therefore
necessary to show that  $\Phi_{T}$ in (\ref{PsiSolution2}) is
gauge invariant. In fact, on applying Stokes theorem to a closed
spacetime path $C$ and using (\ref{relation1}), we find that
$\Phi_T$ changes by
\begin{equation}\label{phase}
  \Delta \Phi_T=\frac{1}{4}\int_\Sigma d\tau^{\mu\nu}J^{\alpha\beta}
  R_{\mu\nu\alpha\beta}\,,
\end{equation}
where $\Sigma$ is a surface bound by $C$ and $J^{\alpha\beta}$ is
the total momentum of the particle. Eq. (\ref{phase}) shows that
(\ref{PsiSolution2}) is gauge invariant and confirms that, to
first order in the gravitational field, the gyro-gravitational
ratio of a Dirac particle is one even when $R=0$. Use of
(\ref{PsiSolution2}) automatically  insures that particle spin and
angular momentum are treated simultaneously and correctly.

\section{Neutrino Geometrical Optics}
\setcounter{equation}{0}

In this Section we study the propagation of a one-flavor Dirac
neutrino in the Lense-Thirring gravitational field \cite{Lense}
represented, in its post-Newtonian form, by
\begin{equation}\label{LTmetric}
 \gamma_{00}=2\phi\,, \quad \gamma_{ij}=2\phi\delta_{ij}\,, \quad
\gamma_{0i}=h_i=\frac{2}{r^3}(\textbf{J}\wedge\textbf{r})_i\,,
 \end{equation}
where
\begin{equation}\label{LTmetric1}
  \phi=-\frac{GM}{r}\,, \quad
  \textbf{h}=\frac{4GMR^{2}\omega}{5r^3}(y,-x,0)\,,
\end{equation}
and  \textit{M}, \textit{R}, $ \mathbf{\omega}=(0,0,\omega)$ and $
\mathbf{J}$ are mass, radius, angular velocity and angular
momentum of the source. This metric is particularly interesting
because it describes a rotating source and has no Newtonian
counterpart. By using the freedom allowed by local Lorentz
transformations, the vierbein field to
$\mathcal{O}(\gamma_{\mu\nu})$ is
\begin{equation}\label{3.5}
 e^0_{\hat{i}}=0\,{,}\quad
 e^0_{\hat{0}}=1-\phi\,{,}\quad
 e^i_{\hat{0}}=h_i\,{,}\quad
 e^l_{\hat{k}}=\left(1+\phi\right)\delta^l_k\,.
 \end{equation}
It is also useful to further isolate the gravitational
contribution in (\ref{3.5}) by writing
 $e^{\mu}_{\hat{\alpha}}\simeq\delta^{\mu}_{\hat{\alpha}}+h^{\mu}_{\hat{\alpha}}$.
 The spin connection can be calculated using (\ref{II.2}) and
(\ref{3.5}) and is
\begin{eqnarray}\label{3.6}
  \Gamma_0 &=& -\frac{1}{2}\phi,_j\sigma^{{\hat 0}{\hat j}}
 -\frac{1}{8}(h_{i,j}-h_{j,i})\sigma^{{\hat i}{\hat j}} \\
  \Gamma_i &=& -\frac{1}{4}(h_{i,j}+h_{j,i})\sigma^{{\hat 0}{\hat j}}
            -\frac{1}{2}\phi,_j\sigma^{{\hat i}{\hat j}}\nonumber\,.
 \end{eqnarray}
Explicit expressions for $\Gamma_\mu$ are given in
Eq.(\ref{spinexpl}) of the Appendix.

 In the geometrical optics
approximation, valid whenever
$|\partial_i\gamma_{\mu\nu}|\ll|k\gamma_{\mu\nu} |$, where $k$ is
the momentum of the particle, the interaction between the angular
momentum of the source and the particle's spin vanishes. This
interaction is quantum mechanical in origin. Then the geometrical
phase $\Phi_G$ is sufficient to reproduce the classical angle of
deflection, as it should, because $\Phi_G$ is common to the
solutions of all wave equations when the spin is neglected.

The deflection angle $\varphi$ is defined by
\begin{equation}\label{tgangle}
  \tan \varphi = \frac{\sqrt{-g_{ij}p_\perp^i
  p_\perp^j}}{p_\parallel}\,,
\end{equation}
where ${\bf p}_\perp$ and $p_\parallel$ are the orthogonal and
parallel components of the momentum with respect to the initial
direction of propagation. In the weak field approximation $\tan
\varphi \simeq \varphi$ and (\ref{tgangle}) reduces to
\begin{equation}\label{angle}
  \varphi \simeq \frac{|{\bf p}_\perp|}{k_\parallel}\,,
\end{equation}
where $k_\parallel = p_\parallel$ is the unperturbed momentum and
$|{\bf p}_\perp|=\sqrt{-\eta_{ij}p^i_\perp p^j_\perp}$, for
$p_\perp^i \sim {\cal O}(\gamma_{\mu\nu})$.

It is clear from (\ref{PhiG}) and (\ref{PsiSolution}) that, once
$\Psi_0(x)$ is chosen to be a plane wave solution of the flat
spacetime Dirac equation, the geometrical phase of a neutrino of
four-momentum $ k^{\mu} $ is given by
\begin{equation}\label{A1}
  \upsilon(x)= -k_\alpha x^\alpha -\Phi_G(x)\,,
\end{equation}
where
\begin{equation}\label{PhiGtilde}
  \Phi_G(x)=-\frac{1}{4}\int_P^xdz^\lambda\left[\gamma_{\alpha\lambda,
  \beta}(z)-\gamma_{\beta\lambda,
  \alpha}(z)\right]((x^{\alpha}-z^{\alpha})k^{\beta}-(x^{\beta}-z^{\beta})k^{\alpha})+
  \frac{1}{2}\int_P^x dz^\lambda\gamma_{\alpha\lambda}k^\alpha\,.
\end{equation}

The components of ${\bf p}_\perp$ can be determined from the
equation

\begin{eqnarray}\label{A2}
  p_i = \frac{\partial\upsilon}{\partial x^i}&=&
  -k_i - \Phi_{G, i}= \\
   & = &  -k_i-\frac{1}{2}\,\gamma_{\alpha i}(x)k^\alpha
   +\frac{1}{2}\, \int_P^x
   dz^\lambda(\gamma_{i\lambda,\beta}(z)-\gamma_{\beta\lambda,i}(z))k^\beta\,.
   \nonumber
\end{eqnarray}

We consider below the two cases of propagation along the $z$-axis,
which is parallel to the angular momentum of the source, and along
the $x$-axis, orthogonal to it. In both instances, the neutrinos
are assumed to be ultrarelativistic, i.e. $dz^0\simeq
dz(1+m^2/2E^2) , E\simeq k(1+m^2/2E^2)$.

\subsection{Propagation along $z$}

Without loss of generality, we consider neutrinos starting from
$z=-\infty$ with impact parameter $b\geq R$ and propagating along
$x=b$, $y=0$. From (\ref{A2}), (\ref{LTmetric}) and
(\ref{LTmetric1}), we find
\begin{eqnarray}\label{p1z}
  p_1 &=& -\frac{1}{2}\left[\int_{-\infty}^zdz^0 \gamma_{00,1}k^0 + \int_{-\infty}^zdz^3 \gamma_{33,1} k^3\right]\\ &=&
  -2k\left(1+\frac{m^2}{2E^2}\right)\int_{-\infty}^z\phi_{,1}dz
   \nonumber\,,\\
  p_2 &=&-\frac{1}{2}\gamma_{02}k^0
  +\frac{1}{2}\int_{-\infty}^zdz^0\gamma_{20,3}k^3=0\,\nonumber\
\end{eqnarray}
and
\begin{eqnarray}\label{pperp}
  (p_{\bot})^1&=&g^{1\mu}p_{\mu}\simeq-p_1=-\frac{2GMk}{b}\left(1+\frac{m^2}{2E^2}\right)\left(1+\frac{z}{r}\right)\,,\\
  (p_{\bot})^2&=&g^{2\mu}p_{\mu}\simeq h_2E=-\frac{4GMR^2\omega
  bk}{5r^3}\left(1+\frac{m^2}{2E^2}\right)\nonumber\,.
\end{eqnarray}

From (\ref{angle}), we finally obtain
\begin{equation}\label{delta0z}
  \varphi =
  \frac{2GM}{b}\left(1+\frac{m^2}{2E^2}\right)\sqrt{\left(1+\frac{z}{r}\right)^2+
  \left(\frac{2R^2b^2\omega}{5r^3}\right)^2}\,,
\end{equation}
which is the deflection predicted by general relativity for
photons, with corrections due to the neutrino mass and the
rotation of the source. In the limit $z\to \infty$ Eq.
(\ref{delta0z}) reduces to
\begin{equation}\label{delta0z1}
 \varphi = \frac{4GM}{b}\left(1+\frac{m^2}{2E^2}\right)\,.
\end{equation}
If, as in \cite{CAR}, we introduce an effective mechanical
momentum
\begin{equation}\label{Peff}
P_{\alpha}^{eff}\equiv
k_{\alpha}+\Phi_{G,\alpha}+\Gamma_{\alpha}\,,
\end{equation}
then there is a residual contribution of the real part of $
\Gamma_{2}$ to $ \varphi $. It is quantum mechanical, it amounts
to $ \frac{3GMR^{2}\omega
bz}{5r^{5}}\left(1-\frac{m^{2}}{E^{2}}\right)$ and is smaller than
all the other terms in $ \varphi $. With this correction, we
obtain
\begin{equation}\label{phi}
 \varphi =
  \frac{2GM}{b}\left(1+\frac{m^2}{E^2}\right)\sqrt{\left(1+\frac{z}{r}\right)^2+
  \left[\frac{2R^2b^2\omega}{5r^3}+\frac{3R^{2}\omega
  z}{5Er^{5}}\left(1-\frac{m^{4}}{E^{4}}\right)\right]^2}\,.
  \end{equation}

\subsection{Propagation along $x$}

In this case neutrinos start from $x=-\infty$ with impact
parameter $b$. For simplicity, we consider neutrinos propagating
in the equatorial plane $z=0$, $y=b$. We find

\begin{eqnarray}\label{p2x}
  p_2 &=&-\frac{1}{2}\gamma_{02}k^0+\frac{1}{2}\int_{-\infty}^xdz^0 \left(\gamma_{20,1}k^1-
  \gamma_{00,2}k^0-\gamma_{10,2}k^1\right)
  -\frac{1}{2} \int_{-\infty}^xdz^1\left(\gamma_{10,2} k^0+\gamma_{11,2}k^1\right)\\ &=&
  -k\left(1+\frac{m^2}{2E^2}\right)\int_{-\infty}^x(2\phi_{,2}+h_{1,2})dx\nonumber\,,\\
  p_3 &=&0 \\\nonumber\
\end{eqnarray}

and
\begin{eqnarray}\label{pperp2}
  (p_{\bot})^2&=&g^{2\mu}p_{\mu}\simeq-p_2+h_2E=-\frac{2GMk}{b}\left(1+\frac{m^2}{2E^2}\right)\left(
  1-\frac{2R^2\omega}{5b}\right)\left(1+\frac{z}{r}\right)\,,\\
  (p_{\bot})^3&=&g^{3\mu}p_{\mu}\simeq-p_3=0\nonumber\,.
\end{eqnarray}

It then follows that
\begin{equation}\label{delta0x}
  \varphi = \frac{2GM}{b}\left(1-\frac{2R^2\omega}{5b}\right)
  \left(1+\frac{m^2}{2E^2}\right)\left(1+\frac{x}{r}\right)\,.
\end{equation}

Contrary to the previous case, the contribution of the angular
momentum of the source does not vanish in the limit $x\to \infty$.
In fact, in this limit we get
\begin{equation}\label{delta0x1}
  \varphi = \frac{4GM}{b}\left(1-\frac{2R^2\omega}{5b}\right)
  \left(1+\frac{m^2}{2E^2}\right)\,,
\end{equation}
which coincides with the prediction of general relativity.
Additional, smaller spin contributions can be obtained from $
\Gamma_{\mu}$ as in Subsection (III.A).

\section{Helicity transitions}
\setcounter{equation}{0}

In what follows, it is convenient to write the left and right
neutrino wave functions in the form
\begin{equation}\label{Psi0}
 \Psi_0 (x) = \nu_{0L,R}e^{-ik_\alpha x^\alpha}=\sqrt{\frac{E+m}{2E}}
  \left(\begin{array}{c}
                \nu_{L, R} \\
                 \frac{{\bbox \sigma}\cdot {\bf k}}{E+m}\, \nu_{L, R} \end{array}\right)
                 \,e^{-ik_\alpha x^\alpha}\,,
\end{equation}
where $\bbox \sigma=(\sigma^1, \sigma^2, \sigma^3)$ represents the
Pauli matrices. $\nu_{L,R}$ are eigenvectors of $\bbox
\sigma\cdot\bbox k$ corresponding to negative and positive
helicity and ${\bar \nu}_{0\, L, R}(k)\equiv \nu_{0\, L,
R}^\dagger (k)\gamma^{\hat 0}$, ${\nu}_{0\, L, R
}^\dagger(k)\nu_{0\, L, R}(k)=1$. This notation already takes into
account the fact that if $ \nu_{\pm}$ are the helicity states,
then we have $ \nu_{L}\simeq \nu_{-}, \, \nu_{R}\simeq \nu_{+}$
for relativistic neutrinos.

In general, the spin precesses during the motion of the neutrino.
This can be seen, for instance, from the contribution $ \Phi_{s}$
in $ \Phi_{T}$. The expectation value of the contribution of $
\Gamma_{0}$ to the effective mechanical momentum (\ref{Peff}) can
in fact be re-written in the form
\begin{equation}\label{Omega1}
 \frac{1}{2}\Psi_{0}^{\dag}
\vec{\Omega}\cdot \vec{\sigma}\Psi_{0},
\end{equation}
where $ \vec{\Omega} \equiv
\frac{GMR^{2}}{5r^{3}}\left(1-\frac{3z^2}{r^2}\right)\vec{\omega}$.
Eq.(\ref{Omega1}) represents the spin-rotation coupling, or
Mashhoon term \cite{mashh}, for the Lense-Thirring metric. Here
rotation is provided by the gravitational source, rather than by
the particles themselves.

We now study the helicity flip of one flavor neutrinos as they
propagate in the gravitational field produced by a rotating mass.
The neutrino state vector can be written as
\begin{equation}\label{neutrinoRL}
 |\psi(\lambda) \rangle = \alpha(\lambda) |\nu_R\rangle
 +\beta(\lambda) |\nu_L\rangle\,,
\end{equation}
where $|\alpha|^2+|\beta|^2=1$ and $\lambda$ is an affine
parameter along the world-line. In order to determine $\alpha $
and $\beta $, we can write Eq. (\ref{PsiSolution}) as

\begin{equation}\label{PsiSolution3}
  |\psi(\lambda)\rangle = {\hat T}(\lambda) |\psi_0(\lambda)\rangle\,,
\end{equation}
where
\begin{equation}\label{Texpression}
  {\hat T}=-\frac{1}{2m}\left(-i\gamma^\mu(x){\cal
  D}_\mu-m\right)e^{-i\Phi_T}\,,
\end{equation}
and $|\psi_0(\lambda)\rangle $ is the corresponding solution in
Minkowski spacetime. The latter can be written as
\begin{equation}\label{psizero}
|\psi_{0}(\lambda)\rangle = e^{-ik\cdot x}
\left[\alpha(0)|\nu_{R}\rangle + \beta(0)|\nu_{L}\rangle
\right]\,.
\end{equation}
Strictly speaking, $ |\psi(\lambda)\rangle $ should also be
normalized. However, it is shown below that $ \alpha(\lambda) $ is
already of $ \mathcal{O}(\gamma_{\mu\nu})$, can only produce
higher order terms and is therefore unnecessary in this
calculation. From (\ref{neutrinoRL}), (\ref{PsiSolution3}) and
(\ref{psizero}) we obtain
\begin{equation}\label{alpha}
\alpha(\lambda)=\langle
\nu_{R}|\psi(\lambda)\rangle=\alpha(0)\langle\nu_{R}|\hat{T}|\nu_{R}\rangle
+ \beta(0)\langle \nu_{R}|\hat{T}|\nu_{L}\rangle\,.
\end{equation}
An equation for $ \beta $ can be derived in an entirely similar
way.

If we consider neutrinos which are created in the left-handed
state, then $|\alpha(0)|^2=0,  |\beta(0)|^{2}=1 $, and we obtain
\begin{equation}\label{PLR}
  P_{L\rightarrow R}=|\alpha(\lambda)|^2=\left|\langle \nu_{R}|\hat{T}|\nu_{L}\rangle\right|^{2}
  =\left|\int_{\lambda_0}^\lambda\langle \nu_R| {\dot x}^\mu \partial_\mu {\hat
  T}|\nu_L\rangle d\lambda\right|^2\\,
\end{equation}
where $\dot{x}^{\mu}=k^{\mu}/m$. As remarked in \cite{CAR}, $
\dot{x}^{\mu}$ need not be a null vector if we assume that the
neutrino moves along an "average" trajectory. We also find, to
lowest order,
\begin{eqnarray}\label{deT}
  \partial_{\mu}\hat{T}&=&\frac{1}{2m}\left(-i2m\Phi_{G,\mu}-i(\gamma^{\hat{\alpha}}k_{\alpha}+m)\Phi_{s,\mu}
  +\gamma^{\hat{\alpha}}(h^{\beta}_{\hat{\alpha},\mu}k_{\beta}+\Phi_{G,\alpha\mu})\right)\\
  \Phi_{s,\lambda}&=&\Gamma_{\lambda}\,,\quad
  \Phi_{G,\alpha\mu}=k_{\beta}\Gamma^{\beta}_{\alpha\mu}\,,\quad
  \nu^{\dagger}_0(\gamma^{\hat{\alpha}}k_{\alpha}+m)=2E\nu^{\dagger}_{0}\gamma^{\hat{0}}\nonumber\,,
\end{eqnarray}
where $\Gamma^{\beta}_{\alpha\mu}$ are the usual Christoffel
symbols, and
\begin{equation}\label{RTL}
 \langle \nu_R |{\dot x}^\mu \partial_\mu {\hat T}|\nu_L \rangle
 =\frac{E}{m}\left[
 -i \, \frac{k^\lambda}{m} {\bar \nu}_R\Gamma_\lambda \nu_L+
 \frac{k^\lambda k_\mu}{2mE}\, (h^\mu_{{\hat \alpha},\,
 \lambda}+\Gamma^\mu_{\alpha\lambda})\nu_R^\dagger \gamma^{\hat
 \alpha}\nu_L\right].
\end{equation}
In what follows, we compute the probability amplitude (\ref{RTL})
for neutrinos propagating along the $z$ and the $x$ directions
explicitly.

\subsection{Propagation along $z$}

For propagation along the $z$-axis, we have $k^0=E$ and $k^3\equiv
k\simeq E(1-m^2/2E^2)$. As in Section III, we choose $y=0,\, x=b
$. With the help of Eq.(\ref{nuGammanu}), we get
\begin{eqnarray}\label{firstcontrib}
  -i\frac{k^\lambda}{m}\bar{\nu}_R\Gamma_\lambda\nu_L&=&\frac{k}{m}\phi_{,1}+i\frac{m}{4E}h_{2,3}\,\,,\\
    \frac{k^\lambda k_\mu
    }{2mE}\, (h^\mu_{{\hat \alpha},\,
    \lambda}+\Gamma^\mu_{\alpha\lambda})\nu_R^\dagger \gamma^{\hat
    \alpha}\nu_L&=&-\frac{k}{2m}\left(1+\frac{k^2}{E^2}\right)\,.\nonumber
\end{eqnarray}
Summing up, and neglecting terms of $ {\cal O}(m/E)^2$,
Eq.(\ref{RTL}) becomes
\begin{equation}\label{RTL1}
\langle\nu_R|\dot{x}^{\mu}\partial_{\mu}\hat{T}|\nu_L\rangle=\frac{1}{2}\phi_{,1}+\frac{i}{4}h_{2,3}\,.
\end{equation}
The contributions to ${\cal O}((E/m)^2)$ vanish. As a consequence
\begin{equation}\label{dadz}
  \frac{d\alpha}{dz}\simeq\frac{m}{E}\frac{d\alpha}{d\lambda}=
  \frac{m}{E}\left(\frac{1}{2}\phi_{,1}+\frac{i}{4}h_{2,3}\right)\,,
\end{equation}
and the probability amplitude for the $ \nu_{L}\rightarrow\nu_{R}$
transition is of ${\cal O}(m/E)$, as expected.

Integrating (\ref{dadz}) from $-\infty$ to $z$, yields
\begin{eqnarray}\label{alfa-z}
  \alpha&\simeq&\frac{m}{E}\left[\frac{1}{2}\int_{-\infty}^zdz\phi_{,1}+\frac{i}{4}h_2(z)\right]\\
  &=&\frac{m}{E}\frac{GM}{2b}\left[1+\frac{z}{r}-i\frac{2\omega
  R^2b^2}{5r^3}\right]\,.\nonumber
\end{eqnarray}
It also follows that
\begin{equation}\label{|alfa-z|}
  P_{L\rightarrow R}(-\infty,z)\simeq \left(\frac{m}{E}\right)^2\left(\frac{GM}{2b}\right)^2
  \left[\left(1+\frac{z}{r}\right)^2+\left(\frac{2\omega b^2R^2}{5r^3}\right)^2\right]\,.
\end{equation}
The first of the two terms in (\ref{|alfa-z|}) comes from the mass
of the gravitational source. The second from the source's angular
momentum and vanishes for $r\rightarrow\infty$ because  the
contribution from $-\infty $ to 0 exactly cancels that from 0 to $
+\infty $. In fact, if we consider neutrinos propagating from 0 to
$ +\infty $, we obtain
\begin{equation}\label{alfa infinito}
  P_{L\rightarrow R}(0,+\infty)\simeq \left(\frac{m}{E}\right)^2\left(\frac{GM}{2b}\right)^2
  \left[1+\left(\frac{2\omega R^2}{5b}\right)^2\right]\,.
\end{equation}
According to semiclassical spin precession equations
\cite{montague}, there should be no spin motion when spin and $
\vec{\omega}$ are parallel as in the present case. This is a hint
that rotation of the source, rather than of the particles, should
produce a similar effect. The probabilities (\ref{|alfa-z|}) and
(\ref{alfa infinito}) mark therefore a departure from expected
results. They are however small of second order. Both expressions
vanish for $ m \rightarrow 0$, as it should for a stationary
metric. In this case, in fact, helicity is conserved \cite{mobed}.
It is interesting to observe that spin precession also occurs when
$ \omega$ vanishes \cite{aldov,casini}. In the case of
(\ref{|alfa-z|}) the mass contribution is larger when $ b<
(r/R)\sqrt{\frac{5r}{2\omega}}$, which, close to the source, with
$b\sim r\sim R$, becomes $ R\omega < 5/2$ and is always satisfied.
In the case described by (\ref{alfa infinito}), the rotational
contribution is larger if $ b/R < 2\omega R/5$ which effectively
restricts the region of dominance to a narrow strip about the $
z$-axis in the equatorial plane, if the source is compact and $
\omega$ is relatively large.

\subsection{Propagation along $x$}

In this case, we put $k^0=E$, $k^1\equiv k\simeq E(1-m^2/2E^2)$.
As in Section III, the calculation can be simplified by assuming
that the motion is in the equatorial plane with $z=0$, $y=b$. We
then have
\begin{eqnarray}\label{contribx}
-i\frac{k^\lambda}{m}\bar{\nu}_R\Gamma_\lambda\nu_L&=&i\frac{k}{m}\phi_{,2}+i\frac{E^2+k^2}{4mE}h_{1,2}-i\frac{E^2-k^2}{4mE}h_{2,1}\,\,,\\
    \frac{k^\lambda k_\mu
    }{2mE}\, (h^\mu_{{\hat \alpha},\,
    \lambda}+\Gamma^\mu_{\alpha\lambda})\nu_R^\dagger \gamma^{\hat
    \alpha}\nu_L&=&-i\frac{k}{2m}\left(1+\frac{k^2}{E^2}\right)\phi_{,2}-i\frac{k^2}{2mE}h_{1,2}\,.\nonumber
\end{eqnarray}
Summing up, and neglecting terms of $ {\cal O}(m/E)^2$,
Eq.(\ref{RTL}) becomes
\begin{equation}\label{RTL1x}
\langle\nu_R|\dot{x}^{\mu}\partial_{\mu}\hat{T}|\nu_L\rangle=\frac{i}{2}\phi_{,2}+\frac{i}{4}(h_{1,2}-h_{2,1})\,.
\end{equation}
The contributions to ${\cal O}((E/m)^2)$ again vanish and we get
\begin{equation}\label{dadx}
\frac{d\alpha}{dx}\simeq\frac{m}{E}\frac{d\alpha}{d\lambda}=\frac{m}{E}\left[\frac{i}{2}
\phi_{,2}+\frac{i}{4}(h_{1,2}-h_{2,1})\right]\sim{\cal
O}(m/E)\,.
\end{equation}
Integrating (\ref{dadx}) from $-\infty$ to $x$, we obtain
\begin{equation}\label{alfax}
  \alpha\simeq i\frac{m}{E}\frac{GM}{2b}\left(1-\frac{2\omega
  R^2}{5b}\right)\left(1+\frac{x}{r}\right)\,
\end{equation}
and
\begin{equation}\label{alfa2x}
  P_{L\rightarrow
  R}(-\infty,x)\simeq\left(\frac{m}{E}\right)^2\left(\frac{GM}{2b}\right)^2\left(1-\frac{2\omega
  R^2}{5b}\right)^2\left(1+\frac{x}{r}\right)^2\,.
\end{equation}
Obviously, the mass contribution is the same as for propagation
along the $z$-axis. However, the two cases differ substantially in
the behavior of the angular momentum term. In this case, in fact,
this term is even, so it does not vanish for $r\rightarrow\infty$.
If we consider neutrinos generated at $x=0$ and propagating to
$x=+\infty$, we find
\begin{equation}\label{pinfinitox}
P_{L\rightarrow
R}(0,+\infty)\simeq\left(\frac{m}{E}\right)^2\left(\frac{GM}{2b}\right)^2\left(1-\frac{2\omega
  R^2}{5b}\right)^2\,.
\end{equation}
The mass term is larger when $ \frac{2\omega R^{2}}{5b}< 1$. At
the poles $ b\sim R$ and the mass term dominates because the
condition $ \omega R <5/2$ is always satisfied. The angular
momentum contribution prevails in proximity of the equatorial
plane. The transition probability vanishes at $ b=2\omega
R^{2}/5$.

\section{Flavor Oscillations}
\setcounter{equation}{0}

Eq. (\ref{PsiSolution2}) can be recast into a form that contains
only first order contributions. By using (\ref{PsiSolution2}) we
find
\begin{equation}\label{4.1}
 \Psi(x)=f(x)e^{-i\Phi_T-ik_\alpha x^\alpha}\Psi_0\,,
\end{equation}
where
\begin{equation}\label{4.2}
  f(x)=\frac{1}{2m}\left[e^\mu_{\hat \alpha}\gamma^{\hat \alpha}(k_\mu+
  \Phi_{G, \, \mu})+m\right]\,
\end{equation}
and $\Psi_0$ now represents the phase independent part of
(\ref{Psi0}) that refers to mass eigenstate neutrinos. The
relationship between flavor eigenstates (Greek indices) and mass
eigenstates (Latin indices) is given by the standard expression
\begin{equation}\label{4.3}
  |\nu_\alpha(x)\rangle = \sum_jU_{\alpha j}(\theta)
  |\nu_j(x)\rangle\,,
\end{equation}
into which (\ref{4.1}) must now be substituted. We find
\begin{equation}\label{4.4}
  |\nu_\alpha(x(\lambda))\rangle = \sum_jU_{\alpha j}(\theta)f_j
  e^{-i{\tilde \Phi}^j_T(x)-ik_\alpha^jx^\alpha}|\nu^j\rangle
  \equiv \sum_jU_{\alpha j}{\hat O}(x(\lambda)) |\nu^j\rangle\,,
\end{equation}
where
\begin{equation}\label{operatoreO}
 {\hat O}(\lambda)=\frac{1}{2m}\left[e^\mu_{\hat \alpha}\gamma^{\hat \alpha}(k_\mu+
  \Phi_{G, \, \mu})+m\right]e^{-i(\Phi_T+k\cdot x)}\,,
\end{equation}
and $U$ is the mixing matrix
 \begin{equation}\label{U}
U =\left(\begin{array}{cc}
                \cos\theta & \sin\theta \\
                 -\sin\theta & \cos\theta \end{array}\right)\,{.}
 \end{equation}
Restricting $\alpha$ to the flavor $e$ for simplicity, we define
the column matrix
\begin{eqnarray}\label{4.5}
  \chi&=&\left(\begin{array}{c}
                \langle \nu_e|\nu_e(x)\rangle \\
                 \langle \nu_\mu|\nu_e(x)\rangle \end{array}\right)
                 =\left(\begin{array}{c}
                \cos^2\theta {\hat O}_{11}+\sin\theta \cos\theta {\hat O}_{12}+
                   \sin\theta \cos\theta {\hat O}_{21}+\sin^2\theta {\hat O}_{22} \\
                                 -\sin\theta \cos\theta {\hat O}_{11}-\sin^2\theta {\hat O}_{12}+
                   \cos^2\theta {\hat O}_{21}+\sin\theta \cos\theta{\hat O}_{22}
                   \end{array}\right)= \\
            &=& U^\dagger \left(\begin{array}{cc}
                 O_{11} & O_{12} \\
                 O_{21} & O_{22} \end{array}\right) U \phi\,,
\end{eqnarray}
where ${\hat O}_{ij}=\langle \nu_i|{\hat
O}(\lambda)|\nu_j\rangle\equiv \nu_i^\dagger {\hat O}\nu_j
=\langle \zeta_i|{\hat O}(\lambda)|\zeta_j\rangle \delta_{ij}$,
$i, j = 1,2$, and $\phi=\left(\begin{array}{c} 1 \\  0
\end{array}\right)$.

The differentiation with respect to the parameter $\lambda$ gives
\begin{equation}\label{4.7}
  i\frac{d\chi}{d\lambda}=U^\dagger\left(\begin{array}{cc}
                 i\frac{dO_{11}}{d\lambda} & 0 \vspace{0.1in} \\
                 0 & i\frac{dO_{22}}{d\lambda} \end{array}\right)U \phi \,.
\end{equation}
By keeping only terms of the first order in
$\gamma_{\alpha\beta}$, observing that $\Phi_{G,\mu}e^\mu_{{\hat
a}, \, \rho}\approx {\cal O}(\gamma_{\mu\nu}^2)$, that $\Phi_{G,
\, \mu}\Gamma_{\rho}\approx {\cal O}(\gamma_{\mu\nu}^2)$ and using
(\ref{PhiGDDerv}) we find that the matrix elements
$idO_{ij}/d\lambda$ in (\ref{4.7}) are of the form
\begin{equation}\label{matrixElemdO}
  i\, \frac{dO_{ij}}{d\lambda}\backsimeq
  A^{(i)}\, O_{ij}+ \sum_k d^{(i)}_{ik}O_{kj} + C^{(i)}_{ij}\,,
\end{equation}
where the index $(i)$ refers to the i-th mass eigenstate, and
 \begin{equation}\label{AdC}
 A^{(i)}={\dot
 x}^{(i)\rho}\left(k_\rho^{(i)}+\Phi_{G,\,\rho}^{(i)}\right)\,,
 \quad
 d^{(i)}_{ij}=\langle \nu_i |{\dot x}^\rho \Gamma_\rho |\nu_j \rangle =
 {\dot x}^{(i) \rho} \langle \zeta_i |\Gamma_\rho |\zeta_j
 \rangle \, \delta_{ij}\,,
 \end{equation}
 \[
  C^{(i)}_{ij}\equiv
  i \frac{{\dot x}^{(i)\rho}}{2m_i}(h_{{\hat \alpha}, \, \rho}^\mu
  k^{(i)}_\mu+\delta_{\hat \alpha}^\mu\Phi_{G, \mu\rho}^{(i)})
  \langle \zeta_i |\gamma^{\hat \alpha} |\zeta_j\rangle \, \delta_{ij}\,
  e^{-i\Phi_G^{(i)}-ik^{(i)}\cdot x}\,.
 \]
The equation of evolution therefore is
\begin{equation}\label{EqEvolution}
 i\frac{d\chi}{d \lambda}
                 = (A_f +d_f) \chi + C \,,
\end{equation}
where
\begin{equation}\label{Af}
  A_f=U^\dagger \left(\begin{array}{cc}
                A^{(1)} & 0 \\
                0 & A^{(2)} \end{array}\right)U\phi\,,
\end{equation}
and
\begin{equation}\label{Cmatrix}
  C= U^\dagger \left(\begin{array}{cc}
                    C_{11} & 0 \vspace{0.1in} \\
                    0 & C_{22} \end{array}\right) U \phi\,.
\end{equation}
A relation similar to (\ref{Af}) links $ d_{f}$ to $ d$.

Up to (\ref{Cmatrix}) the derivation has been general and can be
applied to any metric. It is not possible to say more about the
matrix $ C$ at this point, except that it may contain some
dissipation terms.

In the case of the Lense-Thirring metric, however, $ C $ can be
written as $ C = \bar{C}\chi$. After some algebra, the matrix
elements $ C_{ij}$ become, in fact,
\begin{equation}\label{CijC}
C^{(i)}_{ij}\sim -i \frac{E}{4}\gamma_{00, 3}e^{-i(k^{(i)}\cdot
(x-x_0)-\Phi_G^{(i)})}\delta_{ij}\,.
\end{equation}
Starting from Eq. (\ref{operatoreO}), we can write
\begin{equation}\label{0ij-AB}
  O_{ij} \simeq e^{-ik^{(i)}\cdot (x-x_0)-i\Phi_G^{(i)}}
  (1+A^{(i)}+i2J_0)\delta_{ij}\,,
\end{equation}
where
\begin{eqnarray}\label{Ai}
 A^{(i)} \equiv A &=& \frac{1}{2m_i}(h^\mu_{\hat \alpha}k_\mu^{(i)}+\delta_{\hat
 \alpha}^\mu\Phi_{G, \, \mu}^{(i)})\langle \nu_i|\gamma^{\hat
 \alpha}|\nu_i\rangle
  = \frac{1}{2}h_{\hat 0}^0
  \\
  J_0  &=&
 \frac{GMR^2\omega}{10}\frac{1}{x^2+y^2}\left[\left(\frac{z}{r}\right)^3-\left(\frac{z_0}{r_0}\right)^3\right]
 +
 \frac{GM}{4}\frac{xy}{r}\left(\frac{1}{x^2+z^2}-\frac{1}{y^2+z^2}\right)\,.\nonumber
\end{eqnarray}
To $ {\cal O} (\gamma_{00}m^2/E^2)$\,, $A^{(i)}=A$ is independent
of neutrino mass and energy. It is, in fact a pure geometrical
term. Using (\ref{0ij-AB}), Eq.(\ref{CijC}) becomes
\begin{equation}\label{CijO}
C^{(i)}_{ij}=C_{ij}\sim -i
\frac{E}{4}\frac{\gamma_{00,3}}{1+A+i2J_0}\, {\cal O}_{ij} ={\bar
C}{\cal O}_{ij}\,,
\end{equation}
where
\begin{equation}\label{Cbar}
  {\bar C} \sim  -i \frac{E}{4}\,\gamma_{00,3}\,.
\end{equation}
To ${\cal O}(m^2/E^2)$, $\bar{C}$ does not depend on the mass
eigenstate index $(i)$. Eq. (\ref{matrixElemdO}) therefore becomes
\begin{equation}\label{matrixElemdOCbar}
  i\, \frac{dO_{ij}}{d\lambda}\backsimeq
  A^{(i)}\, O_{ij}+ \sum_k d^{(i)}_{ik}O_{kj} + {\bar C} {\cal
  O}_{ij}\,,
  \end{equation}
and the matrix $C$ is
\begin{equation}\label{Cmatrixbar}
  C= U^\dagger \left(\begin{array}{cc}
                    {\bar C}O_{11} & 0 \vspace{0.1in} \\
                    0 & {\bar C} O_{22} \end{array}\right) U \phi=
                    {\bar C}\chi\,.
\end{equation}
In the equation of evolution (\ref{EqEvolution}) one can then
carry out the transformation $\chi\to
e^{-i\int_{\lambda_0}^{\lambda}{\bar{C}}d\lambda}\chi$ and remove
the ${\bar C}$ dependence by means of a suitable normalization.

The equation of evolution can be re-cast into a more traditional
form by taking ${\dot x}^\rho$ as the tangent vector to the null
world line. Then ${\dot x}^\rho {\dot x}_\rho =0, k^\mu = (E,
k^i)\sim (E, n^iE(1-m^2/2E^2))$, where $n^i$ is a unit vector
parallel to the neutrino three-momentum, $k^0={\dot x}^0$ and
$k^i\approx (1-\varepsilon) {\dot x}^i$, with $\varepsilon \ll 1$.
The matrix elements of (\ref{Af}) contain terms of the form
 $ \dot{x}^{(i)\rho} k^{(i)}_\rho \approx \frac{m_i^2}{2}+\varepsilon
 E^{2}$. The term with $ E$
is diagonal in the matrix of evolution and  does not therefore
contribute to the oscillations. The equation of evolution
(\ref{4.7}) then becomes
   \begin{equation}\label{EqEvolutionFin}
 i\frac{d\chi}{d \lambda}=\left(
           \frac{M_f^2}{2}+\Phi_{G}^{(f)}+\Gamma^{(f)}\right)
            \chi\,,
   \end{equation}
where the flavor mass matrix $M_f$ is related to the vacuum mass
matrix in the flavor base by
 \begin{equation}\label{MassFlavor}
M^2_f=U^\dagger \left(\begin{array}{cc}
                m_1^2 & 0 \\
                 0 & m_2^2 \end{array}\right)U\,,
 \end{equation}
while
 \begin{equation}\label{PhiFlavor}
\Phi_G^f=U^\dagger \left(\begin{array}{cc}
                \langle \zeta_1 | k^\rho\Phi_{G, \rho} | \zeta_1 \rangle  & 0 \\
                 0 & \langle \zeta_2 | k^\rho\Phi_{G, \rho} | \zeta_2 \rangle \end{array}\right)U\,,
 \end{equation}
 and
 \begin{equation}\label{GammaFlavor}
\Gamma^f=U^\dagger \left(\begin{array}{cc}
                \langle \zeta_1 | k^\rho\Gamma_\rho | \zeta_1 \rangle  & 0 \\
                 0 & \langle \zeta_2 | k^\rho\Gamma_\rho | \zeta_2 \rangle
                 \end{array}\right)U\,.
 \end{equation}
To ${\cal O}(\gamma_{\mu\nu}m_i^2/E^2)$, Eq.
(\ref{EqEvolutionFin}) becomes
   \begin{equation}\label{EqEvolutionFinale}
 i\frac{d\chi}{d \lambda}\simeq\left(
           \frac{M_f^2}{2}+k^\rho(\Phi_{G,\,
           \rho}^{(f)}+\Gamma_\rho^{(f)})\right)
            \chi\,.
   \end{equation}
The term in brackets on the r.h.s. of (\ref{EqEvolutionFinale}) is
the effective Hamiltonian of evolution.

By writing the effective mechanical momentum in the form
(\ref{Peff}), squaring it and keeping only first order terms, we
find
\begin{equation}\label{4.9}
  M^2\simeq m^2+2k^\alpha \Phi_{T,\,\alpha}\,.
\end{equation}
and
\begin{equation}\label{4.10}
 |{\bf P}^{eff}|=\sqrt{E^{2}-M^{2}}\simeq E-\frac{M^2}{2E}=E-\frac{1}{E}\left[\frac{m^2}{2E}-k^\alpha\left(\Phi_{G,\,
  \alpha}+\Gamma_{\alpha}\right)\right]\,.
\end{equation}
The third and fourth terms of (\ref{4.10}) represent the
gravitational contributions (\ref{PhiFlavor}) and
(\ref{GammaFlavor}) and provide an explicit expression for the
term $ \vec{p}\cdot \vec{A}_{f}\mathcal{P}_{l}$ in Eq.(27) of
\cite{CAR} when matter effects are neglected. These can be easily
incorporated in the diagonal element of (\ref{EqEvolutionFin}).
The effective potential $V_{\nu_f}$ induced by matter, depends on
neutrino flavors and is defined by \cite{bilenky}
\begin{eqnarray}\label{Vnue}
 V_{\nu_e} &=& -V_{{\bar \nu}_e}=V_0(3Y_e-1+4Y_{\nu_e})\,, \\
 V_{\nu_{\mu,\tau}} &=& -V_{{\bar \nu}_{\mu,\tau}}=V_0(Y_e-1+2Y_{\nu_e})\,,
 \label{Vnuf}
\end{eqnarray}
where $Y_e$ ($Y_{\nu}$) represents the ratio between the number
density of electrons (neutrinos),
\begin{equation}\label{V0}
  V_0=\frac{G_F\rho}{\sqrt{2}m_n}=\frac{\rho}{10^{14}\mbox{gr/cm}^3}\,\,
  3.8\mbox{eV}\,,
\end{equation}
$m_n=938 MeV $ is the nucleon mass, $\rho$ the matter density, and
$G_F$ the Fermi coupling constant.

\subsection{Transition amplitudes}

In order to calculate the transition amplitudes between states of
different flavor, we follow a procedure analogous to that
developed in Section IV for the helicity transitions. To discuss
oscillations, we assume that neutrinos in different mass
eigenstates have the same energy and keep in mind that energy is
conserved in a stationary gravitational field. In order that
different neutrinos interfere at the same final point, we require,
as usual, that the relevant components of the wave function do not
start from the same initial point. This however introduces
negligible corrections in the final results if one assumes, as we
do, that $\Delta m^2=m_2^2-m_1^2\ll m^2$ and $\gamma_{\mu\nu}\ll
1$.

Let us consider, for simplicity, only two flavors and, in
particular, the transition $\nu_e\rightarrow \nu_\mu$. The
transition probability is given by
\begin{equation}\label{probtrans}
  P_{\nu_e\rightarrow
  \nu_\mu}=\frac{|\langle\nu_\mu|\nu_e(\lambda)\rangle|^2}{\langle\nu_e(\lambda)|\nu_e(\lambda)\rangle}\,,
\end{equation}
where $|\nu_e(\lambda)\rangle$ represents at $\lambda$ a neutrino
initially created in the state $|\nu_e\rangle$. From (\ref{4.3})
and (\ref{U}) we obtain
\begin{eqnarray}\label{nunu}
  \langle\nu_\mu|\nu_e(\lambda)\rangle & = &
  \sin\theta\cos\theta\left(\langle\nu_2|\nu_2(\lambda)\rangle-\langle\nu_1|\nu_1(\lambda)\rangle\right)\\
  \langle\nu_e(\lambda)|\nu_e(\lambda)\rangle & = & \cos^2\theta
  \langle\nu_1(\lambda)|\nu_1(\lambda)\rangle+\sin^2\theta\langle\nu_2(\lambda)|\nu_2(\lambda)\rangle\,.\nonumber\
\end{eqnarray}
For a generic mass eigenstate, we find
\begin{equation}\label{ddlambdanu}
  \frac{d}{d\lambda}\langle\nu|\nu(\lambda)\rangle=\left\langle\nu|\frac{d\nu(\lambda)}{d\lambda}\right\rangle=
  -i\dot{x}^\mu k_\mu\langle\nu|\nu(\lambda)\rangle
  +\dot{x}^\mu\langle\nu_0|\partial_\mu\hat{T}|\nu_0\rangle
  e^{-ik_\mu\dot{x}^\mu}\,.
\end{equation}
Since $\langle\nu_0|\partial_\mu\hat{T}|\nu_0\rangle$ is already
$\sim {\cal O}(\gamma_{\mu\nu})$, it is possible to carry out the
substitution
\begin{equation}\label{simeq}
  e^{-ik_\mu\dot{x}^\mu}=\langle\nu|\nu_0(\lambda)\rangle\simeq\langle\nu|\nu(\lambda)\rangle\,.
\end{equation}\
Eq.(\ref{ddlambdanu}) has the first order solution
\begin{equation}\label{solution}
  \langle\nu|\nu(\lambda)\rangle\simeq\exp\left\{\int_P^Q
  d\lambda\dot{x}^\mu\left(-ik_\mu+\langle\nu_0|\partial_\mu\hat{T}|\nu_0\rangle\right)\right\}\,,
\end{equation}
where $P$ and $Q$ are the points at which neutrinos are generated
and detected.

We also obtain
\begin{eqnarray}\label{varnorm}
  \frac{d}{d\lambda}\langle\nu(\lambda)|\nu(\lambda)\rangle & = &
  \left\langle\frac{d\nu(\lambda)}{d\lambda}|\nu(\lambda)\right\rangle+
  \left\langle\nu(\lambda)|\frac{d\nu(\lambda)}{d\lambda}\right\rangle=\\
  & = &
  \dot{x}^\mu\langle\nu_0|\partial_\mu(\hat{T}+\hat{T}^\dagger)|\nu_0\rangle\,,\nonumber
\end{eqnarray}
which has the first order solution
\begin{equation}\label{solnorm}
\langle\nu(\lambda)|\nu(\lambda)\rangle\simeq\exp\left\{\int_P^Qd\lambda\dot{x}^\mu\langle\nu_0|\partial_\mu(\hat{T}+\hat{T}^\dagger)|\nu_0\rangle\right\}\,.
\end{equation}
From (\ref{deT}) we get
\begin{eqnarray}\label{sandwich}
  \langle\nu_0|\partial_\mu\hat{T}|\nu_0\rangle & = & -i\Phi_{G,\mu}-i\frac{E}{m}\bar{\nu}_0\Gamma_\mu\nu_0+
  \frac{k_\nu}{2E}(h^\nu_{\hat{0},\mu}+\Gamma^\nu_{0\mu})\\
  \langle\nu_0|\partial_\mu(\hat{T}+\hat{T}^\dagger)|\nu_0\rangle & =
  &
  \frac{k_\nu}{E}(h^\nu_{\hat{0},\mu}+\Gamma^\nu_{0\mu})\,.\nonumber
\end{eqnarray}
Substituting (\ref{sandwich}) into (\ref{probtrans}), we finally
obtain
\begin{eqnarray}\label{calcprob}
  P_{\nu_e\rightarrow\nu_\mu} &=& \left|\sin\theta\cos\theta\exp\left(-i\int_P^Qd\lambda\dot{x}^\mu
  \frac{E}{m}\bar{\nu}_0\Gamma_\mu\nu_0\right)\right|^2\cdot\\
  &\cdot& \left|\left[\exp\left(-i\int_P^Qd\lambda\dot{x}^\mu(k_\mu^{(2)}+\Phi_{G,\mu}^{(2)})\right)
  -\exp\left(-i\int_P^Qd\lambda\dot{x}^\mu(k_\mu^{(1)}+\Phi_{G,\mu}^{(1)})\right)\right]\right|^2\nonumber\\
  & = &
  \sin^22\theta\sin^2\left(\frac{1}{2}\int_P^Qd\lambda\dot{x}^\mu(\Delta
  k_\mu+\Delta\Phi_{G,\mu})\right)\,.\nonumber
\end{eqnarray}
In deriving (\ref{calcprob}) we have used the fact that the terms
$\bar{\nu_0}\sigma^{\hat{\mu}\hat{\nu}}\nu_0$ are always
proportional to $m/E$, as shown in the Appendix, and that the
contributions from the coefficients $ \Gamma_{\mu}$ are therefore
independent of $m$, $E$ and $k$. It follows from (\ref{calcprob}),
that the effect of the gravitational field is simply obtained by
the substitution $k_\mu\rightarrow k_\mu+\Phi_{G,\mu}\equiv
p_\mu$. This conforms to the semiclassical principle that action
is the quantum phase of the particle.

Eq.(\ref{calcprob}) can be simply rewritten as
\begin{equation}\label{calcprob2}
  P_{\nu_e\rightarrow\nu_\mu}=\sin^22\theta\sin^2\left(\frac{1}{2}\int_P^Qdx^j(\Delta
  k_j+\Delta\Phi_{G,j})\right)\,,
\end{equation}
because the neutrinos have fixed energy, hence $\Delta k_{0} = 0$,
and $\Phi_{G,0}=0$ for static gravitational fields.

If, for simplicity, we consider the motion to be parallel to the
$x^i$-axis, the first term can be written, to $ {\cal O}(m/E)$, in
the traditional form
\begin{equation}\label{oscill}
    \frac{1}{2}\int_P^Qd\dot{x}^j\Delta k_j=\frac{\Delta
    m^2}{4E}(z-z_0)\,.
\end{equation}
Using (\ref{PhiGtilde}), (\ref{A2}) and $ dz^0=(E/k)dz^i$, we can
write
\begin{eqnarray}\label{PhiGi}
\Phi_{G,i}(z)&=&\frac{1}{2}\gamma_{i\alpha}k^\alpha-\frac{1}{2}\int_{z_0}^zdz^\lambda\left(
\gamma_{i\lambda,\beta}-\gamma_{\beta\lambda,i}\right)k^\beta\\
&=&E(2\phi(z)+h_i(z))+\frac{m^4}{4E^3}\phi(z)+E\left(1+\frac{m^2}{2E^2}\right)\phi(z_0)\,,\nonumber
\end{eqnarray}
where, as above, we have retained only the lowest non vanishing
terms in $m/E$. We finally obtain
\begin{equation}\label{deltaphi}
  \Delta\Phi_{G,i}(z)=\frac{m^2\Delta m^2}{2E^3}\phi(z)+\frac{\Delta
  m^2}{2E}\phi(z_0)\,.
\end{equation}
The first term of (\ref{deltaphi}) is the well known gravitational
contribution to the phase of oscillating neutrinos. The second
term corresponds to the redefinition of the constant $E\rightarrow
E\sqrt{g_{00}(z_0)}$ in the non-gravitational term of the
oscillation for a neutrino created at $z_0$, at the desired order
of approximation. With this redefinition, the phase that appears
in (\ref{calcprob}) becomes
\begin{equation}\label{Omega}
  \Omega=\frac{\Delta m^2}{4E^2}(z-z_0)+\frac{m^2\Delta
  m^2}{2E^3}\int_{z_0}^zd\zeta\phi(\zeta)\,.
\end{equation}
The coordinate difference $z-z_0$ is not, however, the physical
distance between two points. This is in fact defined as
$l=\int_{z_0}^z\sqrt{-g_{ii}}dx^i\simeq\int_{z_0}^z(1-\phi)dx^i$,
while $E$ is the neutrino energy measured by an inertial observer
at rest at infinity. Introducing $E_l=E\sqrt{g_{00}(z)}$, which is
the energy measured by a locally inertial observer momentarily at
rest in the gravitational field, the phase $\Omega$ can be
entirely rewritten in terms of physical quantities as
\begin{equation}\label{Omega11}
  \Omega=\frac{\Delta m^2}{4}\int_0^l\frac{dl}{E_l}+\frac{m^2\Delta
  m^2}{2E^3}\int_{z_0}^zd\zeta\phi(\zeta)\,.
\end{equation}
Eq.(\ref{Omega11}) reflects the fact that the curvature of
spacetime affects the oscillation probability through the
gravitational red-shift of the local energy $E_l$ and the proper
distance $dl$.

The oscillation amplitude does not depend on $\gamma_{0i}$. This
means that the relevant oscillation parameters like the phase
$\Omega$, depend quadratically (and \emph{not} linearly) on the
angular momentum of the source. This should be expected because,
as noted by Wudka \cite{wudka}, the quantum mechanical phase is a
scalar, whereas the angular momentum is a pseudovector.

\section{Conclusions}
\setcounter{equation}{0}

In this paper we have applied the weak field solution of
\cite{caipap} to the study of neutrinos in a Lense-Thirring field.
There are advantages to using this solution. It is exact to first
order, is covariant and gauge invariant. It is based on the Dirac
equation and reproduces well all gravitational effects that have
so far been observed \cite{COW,BW}, or have been predicted to
exist for a spin-$ 1/2$ particle \cite{hehlni,lamb}. It also
refers to fermions with unit gyro-gravitational ratio
\cite{dinesh} and thus allows a unified treatment of spin-rotation
and angular momentum-rotation coupling without requiring ad hoc
procedures.

We have applied the solution to the propagation of single flavor
neutrinos. Two cases have been considered in the geometrical
optics approximation. In both instances the larger contribution to
the deviation $ \varphi $ comes from the mass of the source and is
as predicted by Einstein's theory, with corrections due to the
neutrino mass. For propagation parallel to the axis of rotation of
the source, the rotation corrections vanish at infinity. Not so
for propagation perpendicular to the axis of rotation.

It may be argued that spin contributions exist in the
approximation used by introducing an effective neutrino mechanical
momentum given by (III.14) that incorporates a contribution from $
\Gamma_{\mu}$ as in \cite{CAR}. This definition is also supported
by the form of the flavor oscillation Hamiltonian on the r.h.s. of
(\ref{EqEvolutionFinale})-(\ref{4.9}). The contributions that come
from taking the expectation values of $ \Gamma_{\mu}$ are given in
(\ref{phi}) and are much smaller than those of (\ref{delta0z}).

We have then calculated the helicity transition amplitudes of
ultrarelativistic, single flavor neutrinos as they propagate in
the Lense-Thirring field. These transitions are interesting
because at high energies chirality states are predominantly
helicity states and right-handed neutrinos do not interact. The
transition probabilities are of $ {\cal O}( \gamma_{\mu\nu}^{2})$.
Two directions of propagation have again been selected and the
results contain contributions from both mass and angular momentum
of the source. The transitions also occur in the absence of
rotation or with spin parallel to rotation, which is unexpected on
semiclassical grounds. The mass contributions predominate when the
neutrinos propagate from $ r=0$ to $ r = \infty $ (and matter
effects are neglected), provided $ b
>2 \omega R^{2}/5$. There is, however, a narrow region about the
axis of propagation in the equatorial plane where the $ \omega$
contribution is larger. The rotational contribution behaves
differently in the two cases. It vanishes as $ z\rightarrow \mp
\infty$ for propagation along $z$, but not so as $ x\rightarrow
\infty$ in the second case. In addition, when the neutrinos
propagate from $ x=0$ to $ x=\infty$, the mass term dominates in
the neighborhood of the poles, while the contribution of $ \omega$
is larger close to the equator, with no attenuation at $ b=2\omega
R^{2}/5$.

We have also calculated gravity induced, two-flavor oscillations
and derived the relative equation and effective Hamiltonian. A
comparison with the simplified approach of \cite{CAR} leads to
explicit expressions for the term $ \vec{p}\cdot
\vec{A}_{f}\mathcal{P}_{L}$ of these authors. The expressions
contain $ \Gamma_{\mu}$, as expected, and also $ \Phi_{G,\mu}$.
These terms do not exhaust, however, all possibilities if the
metric is not stationary, because of the presence of the matrix $
C$ in (\ref{EqEvolution}). In the general case, therefore, the
evolution of $ \chi $ is more complicated and the effective
Hamiltonian does not comply with the simpler form given in
\cite{CAR}, or in (\ref{EqEvolutionFinale}). Finally, the
transition probabilities do indeed oscillate for the
Lense-Thirring metric, and the curvature of spacetime enters the
oscillation probability through the gravitational red-shift of the
local energy $ E_{l}$ and the proper distance $ dl$.

The neutrino mass appears quadratically in all effects considered
and, in particular, as $ \Delta m^{2}$ in flavor oscillations. It
therefore seems possible, in principle, to obtain the neutrino
masses from an appropriate combination of measurements of these
effects, without resorting to models that might depend on
additional aspects of particle physics. The only ingredient used
is in fact the covariant Dirac equation whose merits have been
extolled at the beginning of this section.

The results presented in this paper agree with those of other
authors, where appropriate. They can be applied to a number of
problems in astroparticle physics and cosmology \cite{dolgov}. For
instance, an interesting question is whether gravity induced
helicity and flavor transitions could effect changes in the ratio
$ \nu_{e}:\nu_{\mu}:\nu_{\tau}$ of the expected fluxes at Earth.

Lepton asymmetry in the Universe \cite{dolgov} also is an
interesting problem. It is known that the active-sterile
oscillation of neutrinos can generate a discrepancy in the
neutrino and antineutrino number densities. The lepton number of a
neutrino of flavor $ f$ is defined by $
L_{f}=(n_{\nu_f}-n_{\bar{\nu}_f})/n_{\gamma}(T)$, where $
n_{\nu_f} (n_{\bar{\nu}_f})$ is the number density of neutrinos
(antineutrinos) and $ n_{\gamma}(T)$ is the number density of
photons at temperature $ T$. As discussed in Sec. IV, the
gravitational field generates transitions from left-handed
(active) neutrinos to right-handed (sterile) neutrinos. If, in
primordial conditions, (\ref{alfa infinito}) and
(\ref{pinfinitox}) become larger, then helicity transitions may
contribute in some measure to lepton asymmetry.

\appendix

\section{Useful Formulae}

\vspace{0.1in}

In this Appendix, we collect useful formulae and results that have
been extensively used in the paper.\\

{\bf Derivatives of $\Phi_G$}. The first derivative with respect
to $x^\mu$ gives
\begin{equation}\label{BPhiGDer}
  \Phi_{G, \mu}=-\frac{1}{2}\int_P^x dz^\lambda (\gamma_{\mu\lambda,
  \beta}-\gamma_{\beta\lambda,
  \mu})k^\beta+\frac{1}{2}\gamma_{\alpha\mu}k^\alpha\,,
\end{equation}
whereas the second derivative is
\begin{equation}\label{PhiGDDerv}
  \Phi_{G, \mu\nu}=k_\alpha \Gamma^\alpha_{\mu\nu}\,,
\end{equation}
where $\Gamma^\alpha_{\mu\nu}$ are the Christoffell symbols of the
second type.\\
 {\bf Christoffel symbols and spin connections}. To
$ \mathcal{O}(\gamma_{\mu\nu})$, the Christoffel symbols for a
Lense-Thirring metric are
\begin{eqnarray}\label{Christ}
  \Gamma^0_{00}&=&0\,,\quad\Gamma^0_{0i}=\phi_{,i}\,,\quad\Gamma^0_{ij}=\frac{1}{2}(h_{i,j}+h_{j,i})\,,\\
  \Gamma^i_{00}&=&\phi_{,i}\,,\quad\Gamma^i_{0j}=\frac{1}{2}(h_{j,i}-h_{i,j})\,,\quad\Gamma^i_{jk}=\delta^j_k\phi_{,i}-\delta^i_j\phi_{,k}-\delta^i_k\phi_{,j}\,.\nonumber
\end{eqnarray}
The spin connection coefficients, already calculated in
(\ref{3.6}), have the explicit form
\begin{eqnarray}\label{spinexpl}
  \Gamma_0&=&-\frac{GM}{2r^3}\left(x\sigma^{\hat{0}\hat{1}}+y\sigma^{\hat{0}\hat{2}}+z\sigma^{\hat{0}\hat{3}}\right)+\frac{GMR^2\omega}{5r^5}\left[(r^2-3z^2)\sigma^{\hat{1}\hat{2}}+3yz\sigma^{\hat{1}\hat{3}}-3xz\sigma^{\hat{2}\hat{3}}\right]\\
  \Gamma_1&=&\frac{3GMR^2\omega}{5r^5}\left[2xy\sigma^{\hat{0}\hat{1}}+(y^2-x^2)\sigma^{\hat{0}\hat{2}}+yz\sigma^{\hat{0}\hat{3}}\right]+\frac{GM}{2r^3}\left(y\sigma^{\hat{1}\hat{2}}+z\sigma^{\hat{1}\hat{3}}\right)\nonumber\\
  \Gamma_2&=&\frac{3GMR^2\omega}{5r^5}\left[(y^2-x^2)\sigma^{\hat{0}\hat{1}}-2xy\sigma^{\hat{0}\hat{2}}-xz\sigma^{\hat{0}\hat{3}}\right]+\frac{GM}{2r^3}\left(-x\sigma^{\hat{1}\hat{2}}+z\sigma^{\hat{2}\hat{3}}\right)\nonumber\\
  \Gamma_3&=&\frac{3GMR^2\omega}{5r^5}\left(yz\sigma^{\hat{0}\hat{1}}-xz\sigma^{\hat{0}\hat{2}}\right)+\frac{GM}{2r^3}\left(x\sigma^{\hat{1}\hat{3}}+y\sigma^{\hat{2}\hat{3}}\right)\,.\nonumber
\end{eqnarray}
{\bf $\gamma$ and $\sigma$ matrices}. We have used the Dirac
representation. The neutrino mass eigenstates are
\begin{equation}\label{A3}
  \nu(x)=\nu_{0L}e^{-ik_\alpha x^\alpha}=\sqrt{\frac{E+m}{2E}}
  \left(\begin{array}{c}
                \nu_{L,R} \\
                 \frac{{\bbox \sigma}\cdot {\bf k}}{E+m}\, \nu_{L,R} \end{array}\right)
                 \,e^{-ik_\alpha x^\alpha}\,.
\end{equation}

\emph{Motion along z direction}. In this case $\nu_L=\left(\begin{array}{c} 0\\
1 \end{array}\right)$ and $\nu_R=\left(\begin{array}{c} 1\\
0 \end{array}\right)$ (i.e. $\nu_{L,R}$ are eigenstates of
$\sigma^3$). To calculate the helicity transitions, we need
\begin{eqnarray}\label{sigmaRL}
\langle \nu_{R}|\gamma^{\hat
0}\sigma^{\hat{0}\hat{1}}|\nu_{L}\rangle &=& -i\frac{k}{E}\,,
  \quad
 \langle \nu_{R}|\gamma^{\hat 0}\sigma^{\hat{0}\hat{2}}|\nu_{L}\rangle =-\frac{k}{E}\,,
 \quad
 \langle \nu_{R}|\gamma^{\hat 0}\sigma^{\hat{0}\hat{3}}|\nu_{L}\rangle
 =0\,,\\
\langle \nu_{R}|\gamma^{\hat
0}\sigma^{\hat{1}\hat{2}}|\nu_{L}\rangle &=&0\,,\quad\quad
 \langle \nu_{R}|\gamma^{\hat 0}\sigma^{\hat{1}\hat{3}}|\nu_{L}\rangle
 =i\,,\quad\quad
  \langle \nu_{R}|\gamma^{\hat 0}\sigma^{\hat{2}\hat{3}}|\nu_{L}\rangle
  =1\,,\nonumber\\
\langle \nu_{R}|\gamma^{\hat 0}|\nu_{L}\rangle &=& \langle
\nu_{R}|\gamma^{\hat 3}|\nu_{L}\rangle=0\,,\quad
 \langle \nu_{R}|\gamma^{\hat 1}|\nu_{L}\rangle=-\frac{k}{E}\,,
 \quad
 \langle \nu_{R}|\gamma^{\hat 2}|\nu_{L}\rangle
 =i\frac{k}{E}\,.\nonumber
\end{eqnarray}

As a consequence of (\ref{sigmaRL}) and (\ref{3.5}) we obtain

\begin{eqnarray}\label{nuGammanu}
  \langle\nu_R|\gamma^{\hat{0}}\Gamma_0|\nu_L\rangle &=&
  \frac{k}{2E}(i\phi_{,1}+\phi_{,2})-\frac{1}{4}[i(h_{1,3}-h_{3,1})+(h_{2,3}-h_{3,2})]\\
  \langle\nu_R|\gamma^{\hat{0}}\Gamma_3|\nu_L\rangle &=&
  \frac{k}{4E}[i(h_{1,3}+h_{3,1})+(h_{2,3}+h_{3,2})]+\frac{1}{2}(i\phi_{,1}+\phi_{,2})\,.\nonumber
\end{eqnarray}

\emph{Motion along x direction}. In this case $\nu_L=\frac{1}{\sqrt{2}}\left(\begin{array}{c} 1\\
-1 \end{array}\right)$ and $\nu_R=\frac{1}{\sqrt{2}}\left(\begin{array}{c} 1\\
1 \end{array}\right)$ (i.e. $\nu_{L,R}$ are eigenstates of
$\sigma^1$). We find
\begin{eqnarray}\label{sigmaRLx}
\langle \nu_{R}|\gamma^{\hat
0}\sigma^{\hat{0}\hat{1}}|\nu_{L}\rangle &=& 0\,,
  \quad
 \langle \nu_{R}|\gamma^{\hat 0}\sigma^{\hat{0}\hat{2}}|\nu_{L}\rangle =\frac{k}{E}\,,
 \quad
 \langle \nu_{R}|\gamma^{\hat 0}\sigma^{\hat{0}\hat{3}}|\nu_{L}\rangle
 =-i\frac{k}{E}\,,\\
\langle \nu_{R}|\gamma^{\hat
0}\sigma^{\hat{1}\hat{2}}|\nu_{L}\rangle &=&1\,,\quad
 \langle \nu_{R}|\gamma^{\hat 0}\sigma^{\hat{1}\hat{3}}|\nu_{L}\rangle
 =-i\,,\quad
  \langle \nu_{R}|\gamma^{\hat 0}\sigma^{\hat{2}\hat{3}}|\nu_{L}\rangle
  =0\,,\nonumber\\
\langle \nu_{R}|\gamma^{\hat 0}|\nu_{L}\rangle &=& \langle
\nu_{R}|\gamma^{\hat 1}|\nu_{L}\rangle=0\,,\quad
 \langle \nu_{R}|\gamma^{\hat 2}|\nu_{L}\rangle=i\frac{k}{E}\,,
 \quad
 \langle \nu_{R}|\gamma^{\hat 3}|\nu_{L}\rangle
 =-\frac{k}{E}\,.\nonumber
\end{eqnarray}

We also obtain (see (\ref{3.5}))

\begin{eqnarray}\label{nuGammanux}
  \langle\nu_R|\gamma^{\hat{0}}\Gamma_0|\nu_L\rangle &=&
  \frac{k}{2E}(-\phi_{,2}+i\phi_{,3})-\frac{1}{4}[(h_{1,2}-h_{2,1})-i(h_{1,3}-h_{3,1})]\\
  \langle\nu_R|\gamma^{\hat{0}}\Gamma_1|\nu_L\rangle &=&
  \frac{k}{4E}[-(h_{1,2}+h_{2,1})+i(h_{1,3}+h_{3,1})]+\frac{1}{2}(-\phi_{,2}+i\phi_{,3})\,.\nonumber
\end{eqnarray}

\acknowledgments Research supported by MURST PRIN 2003.

\end{document}